# Photon absorption and electron scattering by endohedrals

## M. Ya. Amusia [a, b]

[a]) Racah Institute of Physics, The Hebrew University, Jerusalem 91904, Israel

[b]) A. F. Ioffe Physical- Technical Institute, St. Petersburg 194021, Russia

*I dedicate this article to the blessed memory of my friend and colleague professor Walter Greiner*

**Abstract** We concentrate here on photon absorption as well as electron and positron scattering upon endohedrals that consist of a fullerenes shell and an inner atom A. The aim is to understand the effect of fullerene electron shell in formation of corresponding cross-section. We consider the problem substituting the action of a complex multiatomic fullerenes shell by a combination of static pseudopotential and dynamic polarization potential. The electron correlations in the atom A are taken into account in the frame of the random phase approximation with exchange (RPAE). We demonstrate that the fullerenes shell strongly affects the cross-sections, bringing in a number of peculiarities, such as confinement resonances and giant-endohedral resonances and partial wave Ramsauer-type minima. Numerical data are obtained for endohedrals A@$C_{60}$ and A@$C_{60}$@$C_{240}$, where A are noble gas atoms He, Ar and Xe.

## Introduction

The year 1985 has been marked by discovery of a rather exotic in shape multiatomic molecule $C_{60}$ [1]. It presented an almost spherically symmetric construction of 60 carbon atoms with an empty interior. This discovery opened the door for detection of other "empty" molecules, constructed from both carbon and non-carbon atoms. All these objects received the name fullerenes. As to the carbon constructions, in includes now even giants, such as $C_{540}$!

Among other unusual features of fullerenes, one is of particular interest. Namely, it appeared that fullerenes can be "stuffed" by almost any atom A of the Periodic table. One can put inside a fullerene also a small molecule. Inside a big fullerene a small one could be placed also. All such construction received a general name endohedrals and are denoted for atoms as $A@C_N$, presenting an atom A trapped inside a fullerene $C_N$. First endohedral $La@C_{60}$ has been observed one week after the discovery of fullerenes, and presented in [2]

Endohedral is a very complex multi-atomic and many-electron object. Its ab initio calculation is very complicated if possible currently at all. So, in this paper we will use a simplifying approach and simulate the fullerene shell by a spherical potential, adding to it a dynamic polarization potential. Of importance are the electron correlations in atoms that we take into account in the frame of so-called random phase approximation with exchange (RPAE). So, we will treat $A@C_N$ as a "big atom" [3]. In this small review we will present also some results on two-layer endohedrals $A@C_{N1}@C_{N2}$, where fullerene $C_{N1}$ is placed inside $C_{N2}$.

The studies of structure and properties of endohedrals are of interest, since they are scientifically exciting objects, they could exist in Nature and have a whole variety of technological applications. The inner atom A in an endohedral serves as a lamp that illuminates $C_N$ from the inside. As a concrete example, we consider almost spherical $C_{60}$ with a noble gas atom, in most cases, He, Ar and Xe, placed inside. It is essential and simplifying the consideration that noble gas atomic nuclei are located at the center of the fullerene sphere. It is also essential that the fullerene radius $R_F$ is considerably bigger than the atomic radius $R_A$.

The fullerene shell affects the inner atom, modifying its radius and energy levels. The Atom A and fullerene $C_N$ can also exchange electrons, transferring them in both direction and even collectivizing them, totally, or only to some extent. There are good evidences, however, at least for noble gas endohedrals, that these effects are inessential and the inner structure of both objects, $C_N$ and A, are not altered, when one puts A inside $C_N$. However, as we will demonstrate below, $C_N$ strongly affects the processes that took place with participation of A.

We will consider here photoionization, low-energy electron (positron) scattering and decay of vacancies in A, concentrating on the role of $C_N$ upon all these processes. Among the most important effects in this area is distortion of the atomic Giant resonance, formation of Giant endohedral and Interference endohedral resonances [3] and demonstration and analyses of so-called quantum phase additivity in the $e^{\mp}+A@C_N$ scattering.

It deserves to be mentioned also that an “empty” multi-particle construction could in principle be formed from a very big number of nucleons, since in them the Coulomb repulsion is much weaker than the nuclear attraction. It is in place to mention here that the option of a long linear nucleus I have discussed with W. Greiner already in the early ninetieth. One could imagine that an “empty” nucleonic construction could be “staffed” by an ordinary nucleus, at least by a small one.

## Fullerene shell action

The action of $C_N$ includes static action of the fullerene upon atomic A photoelectron or incoming electrons (positrons) in the scattering process $e^{\mp}+A@C_N$. This action is accounted for by introducing pseudopotential $U_F(r)$

$$U_F(r)=\begin{cases}-U_0 & \text{at } R_{in}\le r\le R_{out}=R_{in}+b\\ 0 & \text{at } r<R_{in} \text{ and } R_{out}<r\end{cases}. \quad (1)$$

Here $b$ is the thickness of the fullerene shell that is close to a single-atomic carbon diameter and $R_{in}$ is the inner radius of the fullerene. The concrete values of $U_0$ and $b$ for $C_{60}$ were chosen to reproduce the experimental value of the binding energy of the extra electron in the negative ion $C_{60}^-$ and the low- and medium-energy photoionization cross-section of $C_{60}$ [4].

Fullerenes are polarizable objects. Therefore, an electron that collides with a fullerene shell has to be affected by so-called polarization potential $W(r)$, the simplest form of which is

$$W_F(r)=-\alpha_F\big/2(r^2+d^2)^2 . \quad (2)$$

Here $\alpha_F$ is the fullerene dipole static polarizability and $d$ is the length parameter. In our calculations we put $d=(R_{in}+R_{out})/2\equiv R_F$.

In photoionization of endohedral atoms, potentials (1) and (2) affect the shape of the cross-section, by adding resonance structure that corresponds to reflection of the photoelectron wave by the fullerenes potentials. Maxima in the cross-sections that appear due to the action of potentials (1) and (2) are called confinement resonances.

## Polarization factor

Due to big size and relatively big distance between fullerene nuclei and its electron shell, $C_N$ are highly polarizable objects. This is reflected in its big polarizability. The incoming beam of electromagnetic radiation, in order to ionize the atom A, has to go via the fullerene shell. In dipole approximation that is valid for all photon frequencies in interesting for us energy range, an expression can be derived that connects the electric field $\mathbf{E}_{in}$ inside the fullerene with that of the outside $\mathbf{E}$. To simplify this expression, one has to assume that the radius of fullerene is not simply bigger than the atomic radius, but is much bigger i.e. $R_F / R_A \gg 1$. Applying this inequality, one obtains the following relation (see e.g. in [5]):

$$\mathbf{E}_{in} \equiv \mathbf{E} G_F(\omega) = \mathbf{E}\left[1 - \frac{\alpha_F(\omega)}{R_F^3}\right]. \quad (3)$$

Here $\alpha_F(\omega)$ is the fullerenes dipole dynamic polarizability that at $\omega = 0$ is equal to $\alpha_F$ from (2). The function $G_F(\omega)$ is called polarization factor[1].

It is obvious that if one neglects the potentials (1) and (2), the account of fullerene leads to a very simple relation between photoionization cross section of an endohedral A@$C_N$ - $\sigma^{A@C_N}(\omega)$ and that of an isolated atom A - $\sigma^A(\omega)$:

$$\sigma^{A@C_N}(\omega) = \left|G_F(\omega)\right|^2 \sigma^A(\omega). \quad (4)$$

[1] Here, as in all the rest of this paper the atomic system of units is used, with $e = m_e = \hbar = 1$.

The factor $\left|G_F(\omega)\right|^2$ can have maxima that corresponds to those regions of $\omega$, where $\left|1-\alpha_F(\omega)/R_F^3\right|$ is big. According to (4), this feature leads to a maximum in the photoionization cross-section.

## Destruction and formation of resonances

In this section using concrete examples we demonstrate that fullerene shell can both destroy atomic resonances and form new ones. Let us start with the case of resonance destruction. As a concrete example, let us compare the photoionization cross-section in atom Xe, endohedral Xe@$C_{60}$ and endohedral Xe@$C_{60}$@$C_{240}$ in the area above the ionization threshold of the $4d^{10}$subshell. It is known since relatively long ago that there the cross-section has a high and broad maximum called atomic Giant resonance [6]. Its existence is a manifestation of very strong collective effects in the photoionization of this subshell. In fact, photons in the maximum's frequency region are absorbed by the whole 10-electron 4d-subshell.

The theoretical description of the Giant resonance for atom was achieved by RPAE (see [5] and references therein). Considering photoionization of Xe@$C_{60}$ and two-shell endohedral, e.g. Xe@$C_{60}$@$C_{240}$ [7], one has to take into account that all Xe one-electron excited states are affected by potentials (1) and (2). At $\omega$ bigger than the ionization energy of $4d^{10}$ sub-shell, $G_F(\omega)$ is close to 1. In calculations, we put the radiuses of $C_{60}$ and $C_{240}$ equal to $R_{60}=6.72$ and $R_{240}=13.5$ the depth of the potential wells for $C_{60}$ is $U_0^{60}=0.44$ and for $C_{240}$ is $U_0^{240}=0.53$.

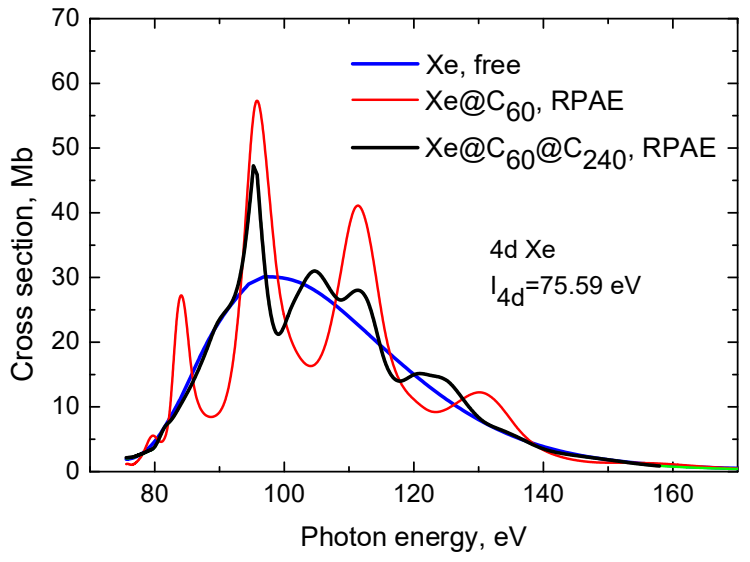

The results of calculations are presented in Fig. 1. We see that the atomic Giant resonance under the action of fullerene shell or shells is destroyed and substituted by a number of narrower and higher resonances.

Fig. 1. Destruction of $4d^{10}$Xe resonance.

Note that the total oscillator strength, i.e. the area under the photoionization curve, of the group of Xe@$C_{60}$ resonances is almost the same as that for $4d^{10}$ Xe itself.

Let us consider photoionization of endohedrals in the region of lower $\omega$. There one has to take into account also the polarization factor $G_F(\omega)$, introduced in (4). The dynamic polarizability $\alpha_F(\omega)$ one can find using the well-known dispersion relation that connects the real and imaginary part of polarizability and takes into account that the imaginary part is simply proportional to the photoionization cross-section of the system under consideration [8]:

$$\mathrm{Im}\,\alpha_F(\omega) = c\sigma_F(\omega)/4\pi\omega,\ \mathrm{Re}\,\alpha_F(\omega) = \frac{c}{2\pi^2}\int_{I_F}^{\infty}\frac{\sigma_F(\omega')d\omega'}{\omega'^2-\omega^2}. \quad (5)$$

Here $c$ is the speed of light. We take $\sigma_F(\omega)$ from experiment. A number of quite accurate measurements exist, starting with published quite long ago in [9]. The dependence of $\alpha_F(\omega)$ upon $\omega$ one can easily understand, having in mind that $\sigma_F(\omega)$ for $C_{60}$ is a powerful broad maximum located at $\omega \approx 22eV$. The so-called total oscillator strength of this maximum is close to 240 – the number of collectivized electrons in $C_{60}$. This has to be compared to the total Xe $4d^{10}$subshell oscillator strength, the biggest for atoms that is equal to about 8. This means that $G_F(\omega)$ factor will be of great importance in this $\omega$ region.

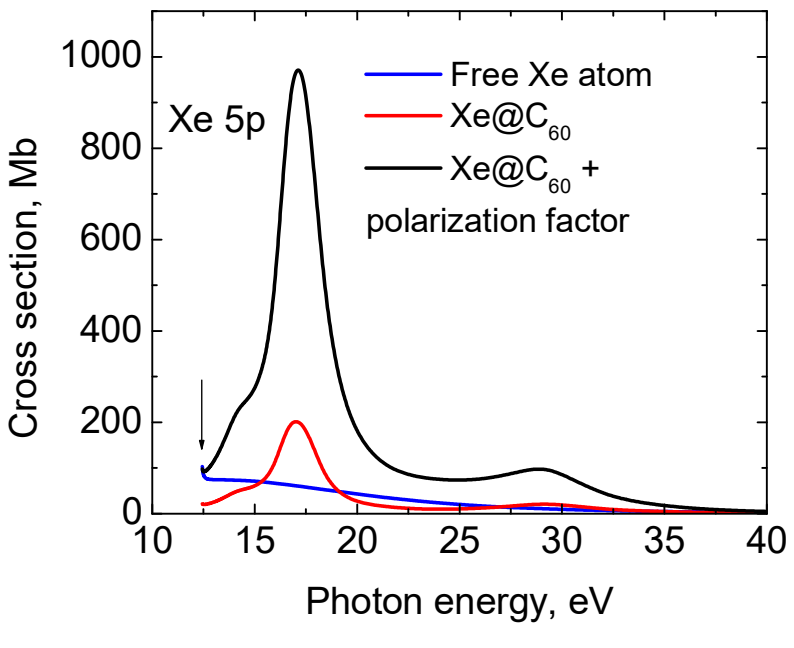

Fig. 2. Endohedral resonance in $5p^6$ of Xe.

Fig. 2 presents the results for photoionization cross-section calculations for outer $5p^6$ subshell of Xe. We include the effects of fullerene shell potentials (1) and (2) as well as polarization factor (3).

The potentials (1) and (2) transform the monotonically decreasing cross section considerably, creating a maximum at 17 ev. But it is the inclusion of the $G_F(\omega)$ factors that leads to dramatic changes. We see that a powerful maximum is formed at about 16 eV and a prominent second maximum appear in the Xe@$C_{60}$ cross-section. The cross-section is more than 900 Mb that is about 45 times bigger than for the isolated atom. We call the respective structure *endohedral resonance* [10]. It is important to note that the total oscillator strength of this resonance is close to 30, that is, by a factor of six more than the corresponding value for the $5p^6$ subshell in the isolated Xe atom. This extra oscillator strength can be "borrowed" only from the oscillator strength of the $C_{60}$. This is an additional evidence of the fact that in the photoionization process in the frequency range of the Xe $5p^6$ subshell, the fullerene and atomic electrons interact very strong, in fact, they became common, to large extent.

## Decay of vacancies

Here we consider the effect of fullerene shell upon the probability of atom's A vacancy decay. Let us start with radiative decay. It can proceed directly by the atom A, emitting a photon. It is also possible that due to interaction between atomic and fullerene electrons the fullerene shell becomes virtually or even really excited and then emits a photon. The amplitudes of two these processes have to be summed determining together the decay probability. We perform calculations assuming as before that $R_F \gg R_A$ [11]:

$$\Gamma_{\gamma,if}^{A@C_N} = \Gamma_{\gamma,if}^{A}\left|1-\frac{\alpha_F(\omega_{if})}{R^3}\right|^2 . \quad (6)$$

Here $\Gamma_{\gamma,if}^{A}$ and $\Gamma_{\gamma,if}^{A@C_N}$ are the radiative width of the vacancy *i* due to its transition to the vacancy *f*; $\omega_{if}$ is the energy of the emitted photon. We see that the effect of fullerene shell upon radiative decay width is determined by the enhancement factor $G_F(\omega)$ that was introduced in (3) in connection to Giant endohedral resonances.

The presence of fullerene shell can open a new, non-radiative or Auger, decay channel. As an example of such a situation, consider the decay of a subvalent vacancy $ns^2$ in a noble gas atom. The transi-tion of an electron from an outer subshell $np^6$ into a vacancy in $ns^2$ leads to emission of a photon with the energy $\omega_{ns,np}$. It cannot decay via emitting another $np$ electron, since the transition has not enough energy to ionize atom A. However, this energy is enough to ionize the fullerenes shell, thus opening an Auger-decay channel and thus increasing by many orders of magnitude the width of a vacancy $ns^2$ in an endohedral, as compared to that in an free atom. In the frame of the same assumptions that leads to (6), one can obtain [11]:

$$\Gamma_{A,if}^{A@C_N} = \Gamma_{\gamma,if}^{A} \frac{3}{8\pi} \left( \frac{c}{\omega_{if}} \right)^4 \frac{\sigma_F(\omega_{if})}{R_F^6}. \quad (7)$$

Here $\Gamma_{A,if}^{A@C_N}$ is the Auger-width of the subvalent vacancy in an endohedral. The ratio $\eta^{A\gamma} \equiv \Gamma_{A,ns,np}^{A@C_N} / \Gamma_{\gamma,ns,np}^{A@C_N}$ varies from 0.5x$10^5$ till 0.5x$10^6$ for noble gas endohedrals from Ne to Xe.

## Electron and positron scattering

At first glance, low-energy elastic scattering cross section of a slow electron should be determined by the size of $C_N$ only, being independent upon the presence or absence of the atom A inside the fullerene. It means that the elastic scattering cross-section $\sigma_{el}(E)$ as a function of incoming electron energy $E$, at $E \to 0$ should be determined only by $R_F^2$, $\sigma_{el}(0) \sim R_F^2$. This should be correct if the low-energy scattering is a classical process.

Direct calculations did not support this assumption [12]. It appeared that the cross-section even at low energies is essentially different from a constant value, depending upon inner structure of the target, namely, upon whether it is an empty fullerene or an endohedral. This difference signals that the low-energy electron (positron) $e^{\mp}$ scattering process is entirely quantum-mechanical. To find the respective scattering phases, one has to solve the following equation for the $l$ partial scattering wave [13]:

$$\left(-\frac{1}{2}\frac{d^2}{dr^2}-\frac{Z}{r}+V_{H_{\mp}F_{-}}(r)+U_{F\mp}(r)+W_{A,\mp}(r)+\right.$$
$$\left.W_{F,\mp}(r)+\Delta W_{FA,\mp}(r)+\frac{l(l+1)}{2r^2}-E\right)P^{A@C_N}_{El,\mp}(r)=0 . \quad (8)$$

Here Z is the atom A nuclear charge, $V_{H_{\mp}F_{-}}$ is the Hartree-Fock potential for $e^-$(Hartree – for $e^+$), $U_{F\mp}(r)$ is the fullerene potential (1) for $e^{\mp}$, $W_{A\mp}(r)$ is the polarization potential of the atom A, and $W_{F\mp}(r)$ is the fullerene polarization potential, determined by (2) for an electron. It proved to be essential to take into account the mutual influence of atomic and fullerenes polarizabilities that we named interference of polarizabilities and contribution of which denoted in (8) as $\Delta W_{FA,\mp}(r)$.

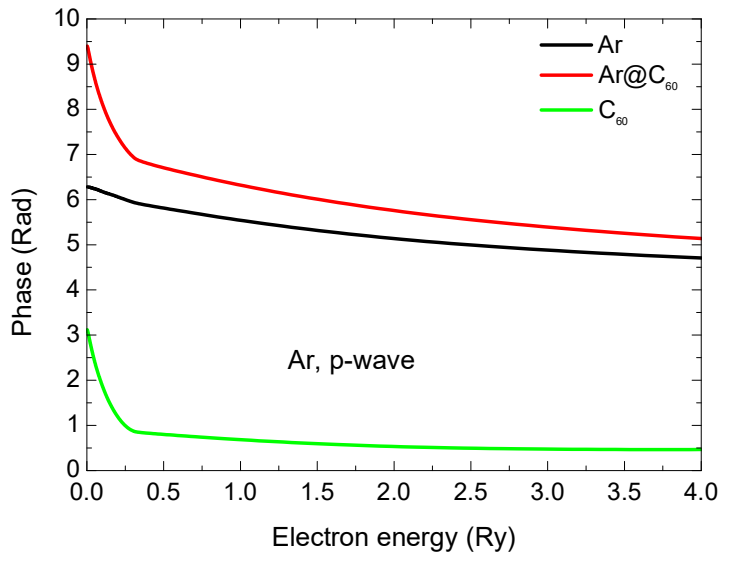

Fig. 3. Rule of additivity of phases in $e^- + Ar@C_{60}$ process.

As is known, the scattering cross-section is expressed via the scattering phases $\delta_{l\mp}(E)$. To simplify the problem, let us at first neglect the interference of polarizabilities, i.e. put $\Delta W_{FA,\mp}(r)=0$. As an example, Fig. 3 presents the elastic scattering $p$ - phase of electrons upon Ar, $C_{60}$, and endohedral $Ar@C_{60}$ [11]. We see that the rule of additivity of phases takes place in this case, namely

$$\delta_l^{A@C_N}(E)=\delta_l^{C_N}(E)+\delta_l^{A}(E) . \quad (9)$$

Note that the polarizability of the atom A is taken into account in the frame of simplified version of RPAE, using the many-body diagram technique (see e.g. [5]). Details on how to calculate $\Delta W_{FA,\mp}(r)$ are presented in [13], but its contribution is important leading to some violation of the phase additivity. We see in Fig. 4 that the electron elastic scattering cross-section by an endohedral that consist of sixty atoms is strongly modified due to presence of a single additional atom inside the fullerenes shell. The presence of an inner Ar leads to Ramsauer-type minima in the s-wave partial cross-section [14].

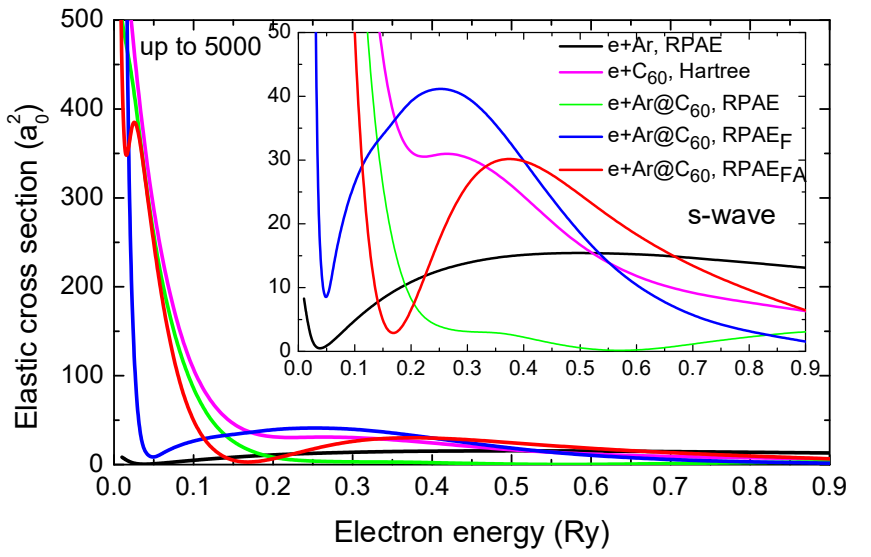

Fig. 4. S-wave contribution to the $e^- + Ar@C_{60}$ cross-section.

At first glance, the scattering of positrons is much simpler to treat than the electron scattering. The only thing what is needed is to neglect the exchange between incoming positron and target electrons. However, the situation is much more complex. Indeed, the incoming positron strongly interacts with virtually excited in the scattering process atomic electrons. They can even form a sort of a temporary bound state called *virtual positronium* $\tilde{P}s$. Its role was recognized in atomic physics long ago [15]. This same effect has to be taken into account in positron – endohedral scattering. Formation of $\tilde{P}s$ modifies the polarization potential. The simplest way to include it is to shift the energy in $\alpha_F(\omega)$ from 0, as it is in (2), to $\omega = I_{Ps}$, where $I_{Ps}$ is the real positronium binding energy.

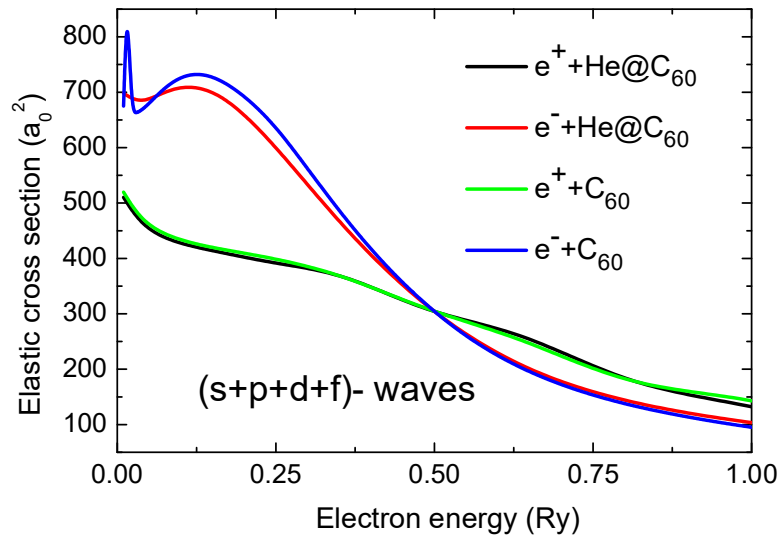

Fig. 5. $e^{\mp}$ scattering upon He and He@C$_{60}$.

We assume also that. $W_{F,+}(r) = -W_{F,-}(r)$. The results of calculations are presented in Fig. 5 [16]. We took into account first four scattering phases, *s*, *p*, *d, f* and present results for $e^{\mp} + C_{60}$ and $e^{\mp} + He@C_{60}$ cross-section, considering them as the sum of all these four contributions.

## Conclusions and perspectives

In this short review we present a number of results in investigation of photon absorption, electron (positron) scattering, and vacancy decay with participation of single- and double-shell endohedrals.

We demonstrate destruction of the atomic Giant resonance and formation, in another photon frequency region, of endohedral Giant resonances due to effect of fullerene shell upon the photoionization cross-section.

We demonstrate that the fullerene shell opens new channels in atomic vacancies decay.

In elastic scattering of slow electrons or positrons upon endohedrals the inner single atom plays, unexpectedly, a prominent role. Due to polarization of the fullerene, the elastic electron scattering cross-sections acquire very big resonances at low energy, and a deep "Ramsauer-type" minimum. The difference between electron and positron cross-sections is very big, the latter being to large extent determined by virtual positronium formation.

We plan to improve the quality of calculations. Expect experimental investigations to check the validity of predictions are certainly desirable.

# Walter Greiner – a colleague and friend.

### (Personal recollections based on half a century long acquaintance)

With the presented article I pay tribute to my old friend professor Walter Greiner, for me Walter, since we were “av du” for more than half a century. Best tribute for a big scientist is discussion of new scientific results. Indeed, each of us is mortal, but Science is immortal.

I vividly remember our first meeting at the Ioffe Institute in former Leningrad, I guess, in the very middle of sixties.

Just at that time, I was pushed off nuclear physics that was my topic before, and have concentrated on applying many-body theory approach to atomic instead of nuclear physics. Walter was interested mainly in nuclear physics, but many-body diagram technique seemed to be good in both domains, namely in studies of multi-nucleon nuclei and multi-electron atoms.

We discussed scientific concrete problems of common interest, such as Giant resonances in both nuclei and atoms. However, particularly great attention we paid to the theory of groups, planning to concentrate in the nearest future on starting experimental activity in this direction of research. It was not mathematics. We discussed the way of organization of big scientific groups that would work not as a simple sum of researches but as a coherent well-organized team. We both agree that scientific productivity without loss of quality of research could be greatly increased in this case. However, it was clear that one has to introduce many organization modifications. For example, Walter suggested regular communication via, perhaps, even written directives from the group leader to his subordinates. Do not know, whether he managed to materialize this idea.

To organize a big and stable group cemented not administratively, but by interest to work under a respected group leader, with members being motivated by love to science, was a challenge. The members of such a group have to be able to sacrifice, at least to some extent, their individuality to the love of collective work. This was a challenge both in the USSR with its kolkhoz traditions and habits, and in the western individualistic community.

We understood that collective work would increase the number of co-authors in any publication. Only by including many co-authors, one could organize a brainstorm that would increase the effectiveness of the research considerably. In such an approach, almost each participant of the brainstorm would become an author. Of course, in case of failure, such an approach could lead to friction among participants and eventually to decay of the group. The organization of collective research with coherent effect belongs to a weakly developed domain of organization of creativity, addressing at first the fundamental issues: whether this is possible at all, and could it be useful. If we come to two, "yes" than one has to find the way "how" implement all this in the real life.

The great experiment of Walter permitted to give very impressive positive answers to all these questions. Walter managed to organize such a group that many people considered as impossible and even counter-productive. I remember numerous discussions between prominent physicist in that time USSR on the subject that was ignited by Walter's visit. Not only I, but also my considerably older and experienced colleagues were very much impressed by his deep intuition as a physicist and the broad range of his ideas. Let me remind you that at that time he was only about thirty years old. My discussions about him with older colleagues, including A. Migdal and Ya. Smorodinsky demonstrated that he was accepted as a promising scientific leader, however all but Migdal were sure that his experiment on new way of organizing scientific research would fail. Fortunately, the big majority proved to be wrong.

With time, Walter managed to build a great pyramid. A number of his former students and group members became prominent scientists. All this is an outstanding achievement.

During our memorable conversations, Walter described interesting formal relations that he planned to have with his students. Twenty years later, I had a privilege almost a year to be a witness of such relations and found that they were useful and fruitful.

During these discussions, Walter has made his prediction of Great Reunification or Great Merge (do not mix it with Great purge!). In spite of its importance, this prediction was never published and even not widely publicized, so, perhaps, I am the only one who can confirm now that indeed such a prediction took place. It was experimentally confirmed almost a quarter of a century later, in 1989. I have in mind the Reunification of Germany. Already during our first discussions, we touch political issues. When I mentioned the now late GDR as an independent German state, Walter answered sharply, with a knock by his palm upon the table: "It can be and will be only one German state that unites all Germans". This was a prophetic prediction of historic process that seemed to me simply impossible giving the power of the USSR and the East block at that time. Let me mention that during this conversation I was looking with fear at the telephone on the table in my room that could inform "interested" what we dare to discuss!

More or less by chance, I have visited Frankfurt in the spring of 1989. My hosts were Greiner and G. Scoff. Walter suggested me to spend in Frankfurt a longer period and nominated for Humboldt research prize. I have received it in 1990 and spent in Germany five months in 1991 and five – in 1992. This prize modified my life powerfully and positively in many aspects. I was very much impressed by the enthusiastic scientific atmosphere that I felt during all my stay. I acquired a number of new friends, and not only among scientists, both Germans and foreigners. This stay was very fruitful for me and I am grateful to Walter for this.

When Walter invited me to come, some of my friends predicted that I would have a hard time since my host will force me to co-authorship. How they were wrong! Walter and I had many discussions, common interests in a number of directions. It appeared that we have almost the same views on a number of scientific (as well as non-scientific) problems, but did not presented even a single common talk or submitted a publication. I did not feel a smallest push toward co-authorship. I attended very many seminars at the Greiner Institute. What I saw on the weekly bases was that in all presented at seminars talks, later published as articles with the speaker as the first author and Walter as the last, in all seen by me cases, he was the main source of the principal idea of the work.

Walter indirectly, by himself as a positive example, taught me how to conduct a seminar to make it a source of inspiration to the speaker. He managed always to concentrate on something interesting and useful in the presented work, finding a reason what for to thank the speaker. His politeness during pallavers, careful praising each speaker, even if the presented work or the talk itself from my point of view did not deserve anything but strong criticism, amazed me. Literally, his motto was “Do not forget to say “thank you” not only to those who are in power and above you, but also to those who depend upon you”. For me it was quite different from what I saw in the USSR, particularly among members of the famous Landau school, and what I followed conducting my own seminar during more then twenty years.

Several years after we met for the first time, he wrote me that moves to atomic physics, having in mind the process of generating of positrons from vacuum under the action of strong, so-called critical, electric field. As far as I know, he was one of the motors, if not the strongest one, behind the idea to create such a field in heavy ion collisions. This was one of the most important ideas implemented in GSI. Although difficulties with separating “overcritical” positrons from that created because of collision process itself proved to be impossible to overcome, the creation of GSI became a great and long lasted success.

During my stay, Walter several times turned to the problem of Jewish Catastrophe under the Nazi regime. He literally felt personal responsibility for the tragedy that happened, and wanted at least somehow to repair the damage not only to Jewish people but also to German science and culture. It was not only words, but also some concrete actions, e.g. organization of long stays and collaboration with Judah Eisenberg, and convincing Walter Meyerhof, son of Otto Meyerhof, Nobel prize winner, to come back to Germany at least for a shot stay. Note that Meyerhof promised never visit Germany after he and his father fled this country after Nazis came to power. Greiner was several times in Israel and became honorary professor of the Tel Aviv University. In 1999 he has a plan to divide the International Symposium *Nuclear Matter-hot and dense*, dedicated to the memory of J. Eisenberg, into two parts, in Tel Aviv and Bethlehem. However, information about planned terrorist attack in Bethlehem prevented materialization of this plan, and all the conference was in Tel Aviv.

Walter was deeply interested in understanding what is going on in Soviet Union during the period that was called perestroika. He took close to his heart the unexpected hardships that almost overnight made the life of Soviet scientist so difficult. He felt not only abstract co-passion but also actively helped in establishing new relations between German and USSR scientists. At first, I want to mention that he managed to bring to Germany for long stays a number of that time young Soviet scientists, to mention only a few of them, M. Gorenshtein, I. Mishustin and my former students A. Soloviev and A. Korol. They were well accepted by the Greiner Institute at the Frankfurt University and after at FIAS that was organized to large extent by Walter.

His help included not only invitations to a number of Soviet scientists to spend in Germany a considerable period. Together, we wrote letters to Riesenhuber, to some other people in the Ministry of science and Education with the aim that could be formulated as "Save Soviet Science". Having in mind much broader cooperation than science, he connected me to the Deutsche Bank Vice-President.

Walter's kindness and attentiveness had no limits and not only I, but a number of people of the former USSR are grateful to him forever.